# Analyzing the Overhead of Filesystem Protection Using Linux Security Modules

Wenhui Zhang, Trent Jaeger, Peng Liu
wenhui@gwmail.gwu.edu
Penn State University
State College, P.A.

## ABSTRACT

Over the years, the complexity of the Linux Security Module (LSM) is keeping increasing (e.g. 10,684 LOC in Linux v2.6.0 vs. 64,018 LOC in v5.3), and the count of the authorization hooks is nearly doubled (e.g. 122 hooks in v2.6.0 vs. 224 hooks in v5.3). In addition, the computer industry has seen tremendous advancement in hardware (e.g., memory and processor frequency) in the past decade. These make the previous evaluation on LSM, which was done 18 years ago, less relevant nowadays. It is important to provide up-to-date measurement results of LSM for system practitioners so that they can make prudent trade-offs between security and performance.

This work evaluates the overhead of LSM for file accesses on Linux v5.3.0. We build a performance evaluation framework for LSM. It has two parts, an extension of LMBench2.5 to evaluate the overhead of file operations for different security modules, and a security module with tunable latency for policy enforcement to study the impact of the latency of policy enforcement on the end-to-end latency of file operations.

In our evaluation, we find opening a file would see about 87% (Linux v5.3) performance drop when the kernel is integrated with SELinux hooks (policy enforcement disabled) than without, while the figure was 27% (Linux v2.4.2). We found that performance of the above downgrade is affected by two parts, policy enforcement and hook placement. To further investigate the impact of policy enforcement and hook placement respectively, we build a Policy Testing Module, which reuses hook placements of LSM, while alternating latency of policy enforcement. With this module, we are able to quantitatively estimate the impact of the latency of policy enforcement on the end-to-end latency of file operations by using a multiple linear regression model and count policy authorization frequencies for each syscall. We then discuss and justify the evaluation results with static analysis on our syscalls' call graphs, which is call string analysis enhanced with execution order analysis.

## 1 INTRODUCTION

The trade-off between security and performance is an important consideration in the design authorization systems to enforce security policies in filesystems. During the past decade, we observe that the relative overhead of authorization hooks in Linux systems has been increasing substantially. As shown in Table 1, for example, from Linux v2.4.2 to v5.3.0, the relative overhead of open increases from 27% to 87%. As performance overhead has long been a serious concern of filesystem developers, a thorough syscall-level measurement study on the impact of authorization hooks on filesystem performance is highly desired. To make such a measurement study rigorous and thorough, we believe that the following four basic requirements must be met. (1) The impact of the **placement** (i.e., where to place a hook) aspect of authorization hooks and that of the **policy enforcement** aspect (i.e., to see if an access violates the security policy) should be measured separately. This decoupling is important for us to figure out which aspect is a dominant reason.
(2) The measurement study should be comprehensive. That is, every widely-used system all should be taken into consideration. (3) The measurements should be precisely measured, libc calls etc. in user-space tests could result in misleading measurements. (4) Depth test should be conducted to syscalls related to directorial accesses. The previous measurement studies, such as LSM (v2.5.15) [24] and SELinux (Linux v2.4.2)[19, 38], fall short of meeting these four requirements.

Besides this observation, we are motivated to revisit the overhead of LSM implementations due to three reasons. First, the size of the kernel code and LSM hook continues to grow. On Linux v5.3.0, there are 18,883,646 LOC and 224 hooks, while there were 3,721,347 LOC and 122 hooks in Linux v2.6.0. Second, new features are introduced into the kernel monthly. Flexible module stacking is a feature introduced to LSM in year 2019 [37] and integrity protection of security attributes was introduced to LSM [46] in 2008. The performance impact of these features has not been evaluated before. Thirdly, various security modules (e.g. AppArmor, TOMOYO and SMACK) are merged into mainstream, they implement difference set of hooks. The performance impact of implementing different set of hooks has not been evaluated before. These three reasons make previous results less relevant to research investigations on LSM.

In this work, we provide a systematic evaluation of overhead of LSM hooks on file accesses. LSM introduces hundreds of security checks (224 in v5.3.0) scattered over 18 million LOC kernel code (v5.3.0). To meet the aforementioned requirements in measuring the performance impact of the hooks is challenging work due to the complexity of the code. The hooks that each security module chooses to implement vary greatly even for the same kernel object. For example, SELinux implements 31 inode-based hooks, and AppArmor implements 1, while SMACK implements 22. To evaluate the performance impact of the hooks, we need to decouple the interfaces from the other functionalities implemented in the reference monitor system. To do this, we disable the policy enforcement code, which is implemented for querying policy from policy store and policy parsing, processing and checking, in the LSM-based security modules. By doing this, the impact of invoking the hooks is not shadowed by the other parts of LSM. We evaluate the overhead of the hooks in four major LSM-based security modules which

**Table 1: Performance Differs as LSM and Hardware Evolves. Latency is evaluated with default setting of LMBench2.5 for open, stat and creat, the lower, the better. Throughput of copy, read and write is evaluated with 4KB files, the higher the better. Latency of read, write and copy is evaluated with 0KB files, the lower the better. Overhead is compared to the kernel with pure DAC protection, the lower, the better.**

| Paper | Version | Hooks | CPU | L2 Cache | Memory Size | Storage | open | stat | creat | copy | read | write |
|---|---|---|---|---|---|---|---|---|---|---|---|---|
| LSM [24]* | 2.5.15 | 29 | 700MHz | 2000KB | 1GB | SCSI disk | 7.13□s | 5.49□s | 73□s | 191MB/s | 368MB/s | 197MB/s |
| Overhead: | | | | | | | 2.7% | 0% | 0% | 0% | 0% | 0% |
| Current LSM‡ | 5.3.0 | 224 | 2.50 GHz | 3072KB | 8GB | SSD | 1.5□s | 0.8□s | 13□s | 2.45GB/s | 10.34G/s | 4.96GB/s |
| Overhead: | | | | | | | 7.5% | 1.3% | 5.1% | 3.6% | 5.5% | 0.6% |
| SELinux [19]† | 2.4.2 | 122 | 333MHz | 512KB | 128MB | N/A | 14□s | 10.3□s | 26□s | 21□s | N/A | N/A |
| Overhead: | | | | | | | 27% | 28% | 18% | 10% | N/A | N/A |
| Current SELinux‡ | 5.3.0 | 204 | 2.50 GHz | 3072KB | 8GB | SSD | 2.2□s | 1.1□s | 18.5□s | 0.7□s | 0.36□s | 0.37us |
| Overhead: | | | | | | | 87% | 30% | 15.9% | 10.5% | 13.2% | 3.8% |

* This is carried out using LMBench2.5 executed on a 700 MHz Pentium Xeon computer with 1 GB of RAM and an ultra-wide SCSI disk.
† This is carried out using LMBench2.5 executed on a 333MHz Pentium II with 128M RAM.
‡ This is tested with LMBench2.5 on 6th Generation Intel® Core™ i7-6500U 2.50 GHz processor with 2 cores, at 1,442MHz each

are SELinux, AppArmor, Smack and TOMOYO. Results show that different security modules incur different overhead.

We further evaluate performance impact of module stacking of the LSM framework. Module stacking has been introduced into the LSM framework lately. It allows the system to have more than one active security module. With module stacking, the system follows a whitelist-based checking order. For example, capabilities modules could be stacked on top of one of the other major modules, or vice versa. We find that different stacking orders have different performance impact.

To ensure the property of being tamper-proof, the LSM framework uses integrity modules for measuring and verifying integrity of a file (i.e., an inode) and integrity of metadata associated with it. There are 12 hooks (Linux v5.3.0) in LSM which have been instrumented with integrity protection code. Such code also impacts the performance of the hooks. Integrity module supports various integrity measurements, such as auditing, Integrity Measurement Architecture (IMA), Linux Extended Verification Module (EVM). We evaluate performance overhead of auditing, IMA and EVM.

Last but not least, to further investigate where the performance downgrade is coming from, we implement a special-purpose Linux security module to study the relationship between the latency of policy checking and the end-to-end latency of file accessing system calls. We control the latency of policy checking in our security modules and measure the end-to-end latency of system calls. We find for most system calls, the relationship is linear; also, for certain system calls, such as open and stat, the linear coefficient is proportional to the number of components in the input path. This suggests that caching the policy-checking results for directories can improve the performance of meta-data accessing for the file and sub-directory underneath them.

In summary, in this work we make the contributions as follows:

- The overhead of a LSM-based security module is caused not only by invoking the hooks but also by policy enforcement. Prior work only measured the **combined** overhead. In this work, we measure the overhead caused by invoking the hooks (i.e. hooking) and the overhead caused by policy enforcement **separately**. We compare hooking overheads of a spectrum of LSM-based security modules. We also evaluate stacking order's impact on performance overhead of these LSM-based security modules. We find that stacking orders can make overhead increase to 45x for TOMOYO and 61% for SELinux. We evaluate the performance impact of integrity measurements (i.e., auditing, IMA and EVM) on SELinux.
- We decouple policy enforcement and hook placement, and implement a special-purpose Linux security module to study the relationship between the latency of policy checking and the end-to-end latency of system calls for file accesses. By using a multiple linear regression model, we quantify the impact of the latency of policy enforcement on the end-to-end latency of file operations, and identify the over-worked permission checks on Linux VFS.
- We discuss and identify the causes of the above-measured overhead, together with static analysis of syscall call graphs for justification of our findings.

The rest of the paper is organized as follows. Section 2 describes background knowledge for LSM. Section 4 explains methodology we used to drive our analysis. We summarize our main findings in Section 5 before zooming into performance overhead root causes discussion in Section 6. Section 7 reviews previous evaluation works. Section 8 concludes the work.

## 2 BACKGROUND

In this section, we present background knowledge of evolution of hooking overhead in LSM. We explain execution path of access control during accessing files, integrity protection of LSM and the mechanism of stacking multiple LSM security modules. Lastly, we discuss limitations of LMBench on evaluation of LSM.

**Evolution of Hooking Overhead in LSM.** LSM framework is introduced in 2002 [42], which supports an interface that allows Linux to have mandatory access controls. It is firstly merged in Linux v2.5.29, with 29 hooks and 1,249 LOC. The hook number and implementation becomes more and more complex since then. SELinux [19, 24, 38], the first mandatory access control system in mainline Linux, is incorporated into the Linux v2.6.0, with 122 hooks and 10,684 LOC. Increased LSM adapted enhancements aimed at improving performance [23], such as hooking on network flow, rather than packets [15]. Smack [6, 35] is adopted to LSM since Linux v2.6.25, TOMOYO [13] is merged into Linux v2.6.30, and AppArmor [2] into v2.6.36. LSM has been supporting more and more MAC since then, when Linux v4.18.5 releases (Ubuntu 16.04), 190 LSM hooks are defined. Now, Linux v5.3.0 (Ubuntu 18.04) has 224 hooks (65,793 LOC), with 204 for SELinux, 68 for AppArmor, 108 for SMACK, 28 for TOMOYO. As the the number of hooks grows, it becomes complex to reason the root causes of LSM's overhead.

**Entangled Code for Filesystem Access Control.** Theoretically, access control in Linux includes two parts: (1) DAC and (2) MAC. DAC is a must for access control in Linux, while MAC coexists as a supplementary since Linux version 2.6 [24, 42]. The architecture of DAC-MAC coexistence is shown in Figure 1. The workflow of Linux's access control is as follows. User space programs work with external resources via the kernel, and make requests for accesses through system calls. When a program executes a system call to access files, for example, open a file, the kernel performs a number of checks. Linux first verifies the correctness of the passed arguments, checks the possibility of their allocation. If the file exists, the request will be handed over to kernel functions. Kernel functions check if the program has the permission to obtain the requested resource by DAC, through UID, GID and modes (i.e., read, write, execute) validation. If the request passes DAC, it is handed over to LSM. The LSM hooks handle these requests, and query LSM-based security modules (e.g. SELinux) for permissions. For example, function *inode_permission* (i.e., in file fs/namei.c) first checks for read permission on the filesystem itself (i.e., *sb_permission* in fs/namei.c). Then it calls *inode_permission* (i.e., in file fs/namei.c) to check for read/write/execute permissions. Afterwards it checks POSIX ACL on an inode through *do_inode_permission* (i.e., in file fs/namei.c). This procedure concludes DAC permission checking. Finally, LSM related permissions (e.g. SELinux) are checked through calling *secu- rity_inode_permission* (i.e., in file security/security.c). However, the implementation of DAC and MAC is not always cleanly separated. **Hooking and Reference Monitor Concept.** Reference Mon- itor Concept has three requirements: (1) Complete Mediation, (2) Tamper-proofing, and (3) Verifiability. This paper investigates into overhead of reference monitor systems, in particular LSM-based security modules, from the above three aspects. Complete media- tion requires mediating all security-sensitive operations through security hooks.

Hooks are placed on the execution path of security- sensitive operations, which handle shared security-sensitive objects (SSOs), and they introduce overhead to these security-sensitive op- erations. LSM-based security modules implement distinct subsets of security hooks, also stacking of LSM-based security modules introduce overhead as well. In this paper, we evaluate the over- head for distinct security modules by evaluating the performance of the subsets of hooks they each implement. We also evaluate the performance impact of different stacking orders. Tamper-proofing requires that module-defined protection state, e.g., module-defined labels of processes, and files are protected. For example, the In- tegrity Measurement module protects the security attributes and security blobs of files from being modified by malicious processes through auditing, IMA and EVM. In this work, we also evaluate the overhead introduced by integrity measurement through the above 3 aspects. Verifiability requires the policies of the authorization mechanism to be verified to enforce the expected security goals.

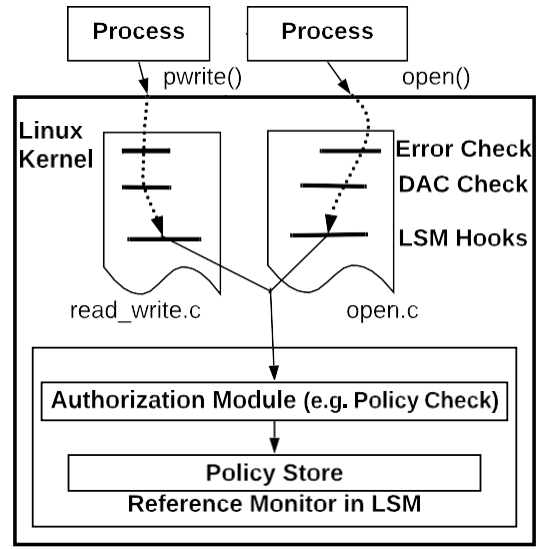

**Figure 1: Linux Security Modules Framework.**

Distinct LSM-based security modules often perform authorization using different policy models, creating module-specific policy se- mantics. However, the impact of the policy model on overhead is less significant than the costs related to complete mediation (i.e, hooking) and tamper-proofing defenses. Regardless of the policy model all have to perform a similar authorization check. This paper focuses on the mediation and the checking and each's overhead for that, integrity measurement overhead (for tamper-proofing re- quirement) is also investigated. The rest of the reference monitor guarantees are provided by the kernel and the policy configuration, which is out of scope.

**Integrity Protection of Security Attributes in LSM.** LSM utilizes a security-tag system, such as extended attributes in Ext4, BtrFS and etc., to enforce security. Integrity module uses 12 hooks (Linux v5.3.0) to collect, appraise and store security attributes (i.e., , integrity xattr value) for operations. It measures and verifies the integrity xattr and provide protection of security attributes for LSM. Integrity module supports different integrity measurements, such as auditing, the Integrity Measurement Architecture (IMA) [33] and the Extended Verification Module (EVM) [12]. Auditing keeps track of the pointers to the security_operations, and records attempts at access according to syscall audit rules. IMA keeps track of hashes of the files. Each newly calculated file hash extends one of the Platform Configuration Registers (PCRs). The value stored in the PCR is the aggregated hash of all files seen by IMA. EVM is designed to detect when files on disk have been modified while the system was shut down. It creates HMAC out of different types of metadata of individual files including security related extended attributes, file owner and group and etc.

**Module Stacking in LSM.** Flexible LSM stacking [6] has been introduced to LSM framework lately. It allows more than one LSM modules to be active in the system. It is useful in the containerized environment where the container requires a different LSM module to what the host enables [36]. An example is to run Ubuntu con- tainers on a host with RedHat distribution [36]. The former needs

Table 2: Lines per Hook Varies (Linux version 5.3).

| Name | # of Hooks | General Hooks | LOC | LOC/Hook |
|---|---|---|---|---|
| **capabilities** | 18 | 18 | 767 | 43 |
| **SELinux** | 204 | 170 | 21266 | 104 |
| **AppArmor** | 68 | 62 | 11918 | 175 |
| **SMACK** | 108 | 100 | 5369 | 50 |
| **TOMOYO** | 28 | 27 | 8245 | 295 |
| *Integrity\** | 12 (5/7) | 11 (5/6) | 6107 | 509 |
| LSM | 224 | 153 | 65793 | N/A |

AppArmor while the later only enables SELinux by default. In this scenario, the host needs both SELinux and AppArmor to be active. When multiple LSM modules are active in the system, the order in which checks are made is specified by CONFIG_LSM during the compile time. The checking follows a white-list mechanism, which only gives access to objects if all security modules approve. If the access is not granted by one security module, it will not be checked by the next security module. Without a specific LSM built into the kernel, the default LSM will be the Linux capabilities system. Most LSM-based security modules choose to extend the capabilities system, building their checks on top of the defined capability hooks. For more details on capabilities, see capabilities(7) in the Linux man-pages project.

**LMBench tests on Filesystem syscalls.** LMBench builds userspace tests for filesystem operations involving one or more syscalls and measures the latency and/or throughput of these syscalls. However, LMBench is not designed for evaluating individual syscalls, nor does LMBench span all filesystem syscalls. Among the 382 system calls in Linux version 5.3.0, 43 of them perform file access operations, of which POSIX defines a minimum set of operations that must provided for file access [11, 28, 41]. Many of these filesystem syscalls access *security-sensitive objects* (SSO) [15], such as superblock, path, inode, dentry and file data structures, requiring authorization of access to those data structures by invoking LSM hooks. In addition, many syscalls must be performed atomically to maintain correctness under concurrent access. However, LMBench also does not cover all filesystem atomic functions. Thus, LMBench is not directly applicable for measuring overhead for LSM operations.

## 3 OVERVIEW

In this paper, we analyze the overhead (on filesystems) caused by hook placement and policy enforcement. One objective of this measurement study is to decouple these two factors, so that the influence of each factor can be separately measured and analyzed. Another objective of this measurement study is to identify the causes of the measured overheads.

To achieve these two objectives, a challenge is that LSM interface and LSM-based security modules' implementations are complex. Different security modules provide implementations with different sets of hooks. As is shown in Table 2, for example, SELinux (Linux v5.3) implements 10 hooks on files, 31 hooks on inodes, 2 hooks on dentries, 13 hooks on superblocks; while AppArmor (Linux v5.3) implements 1 hook on inodes, 7 hooks on files, 3 hooks on superblocks, and 10 hooks on file paths. In addition, the complexity of the implementations (based on lines of code) for each hook varies, see Table 2. Capabilities and TOMOYO on average have 43 LOC per hook and 295 LOC per hook, respectively. SELinux and SMACK both implement security_file_permission. However, the number of lines of code they use for implementing this hook are different. Furthermore, the stacking feature adds more complexity for analyzing overhead of filesystem protection of LSM-based security modules.

In order to identify performance bottlenecks in LSM implementations, we build a test suite for VFS syscalls to measure latency and throughput (i.e. operations per second) tests. In addition, we have developed a special purpose, latency-controllable security module to diagnose the performance impact of policy enforcement (e.g. authorization through security_inode_permission and security_file_permission) and its impact on the performance of VFS syscalls.

**Scope of this work.** This measurement study is for filesystem developers, not for application developers. In our view, diagnosing the performance bottlenecks at the syscall level is to a large extent orthogonal to diagnosing the bottlenecks at the application level. Therefore, an application-level measurement study is out of scope.

## 4 METHODOLOGY

In this section, the methodology of our evaluation is explained. The overhead imposed by LSM is a composite of the overhead of hook placement (i.e., the number of hooks invoked) and the policy enforcement overhead (i.e., policy authorization). This work focuses on evaluating how policy enforcement performance impacts the overhead of file operations. We would like to study the overhead of hooks' implementations of policy enforcement for file operations in LSMs and how different pathname-patterns impact hook invocation frequency and the end-to-end performance of file operations. Thus, a evaluation framework is built with below two parts: (1) an extension of LMBench2.5 that we call **LMBench-Hook** that tests 14 filesystem syscalls and (2) a **tunable security module** that enable us to control the policy enforcement latency for assessing the impact of policy enforcement overhead. We use LMBench-Hook to comparatively measure the overhead of a variety of hooking configurations determined by the hooks they support, the LSM stacking orders, and uses of integrity measurement. A tunable security module is further developed to study how the latency of policy enforcement impacts the end-to-end performance of file accesses. At the end, we discuss the limitations of our evaluation framework.

### 4.1 Extending LMBench

Previously, the authors of [42] and [19] used LMBench2.5 [22] to evaluate performance impact of hooking for LSM and SELinux, respectively, see Table 1. They evaluated open, stat and creat for a particular directory/file. They only tested a subset of filesystem operations. Firstly, filesystem operations is more than open, stat and creat. Hooks are also placed on system calls, such as read/write/copy, link/unlink/symlink, chmod etc. Secondly, some filesystem operations' performance are influenced by directory depth, such as open and stat. In this work, we would like to evaluate other hooks invoked by filesystem operations, which further include read/write/copy, link/unlink/ symlink, chmod, rmmdir/mkdir, and etc. We test system call open and stat's performance with varying path name lengths. To meet our evaluation purpose, we extend LMBench2.5 as LMBench-Hook, to measure the performance impact of hooking on file accesses. We modify LMBench2.5 code to execute tests over more syscall types and to enable control over the input paths for the tests that need a path name. Apart from the changes of configuring input paths, we reuse LMBench2.5's code to measure the latency of the file operations listed in Table 3 expect rmdir, mkdir, read, write and copy. For these five file operations, we add new tests and also measure their throughput (operations per second) instead of latency.

**Table 3: The List of the Benchmarks of LMBench-Hook, the System Calls Invoked by them in Order and their Category.**

| No. | Test Name | Syscall Name | Class |
|---|---|---|---|
| 1 | **open** | open, close | File Ops |
| 2 | **openat** | openat, close | File Ops |
| 3 | **rename** | rename | File Ops |
| 4 | **creat** | rename, creat, close | File Ops |
| 5 | **mkdir** | mkdir | Dir Ops |
| 6 | **rmdir** | rmdir | Dir Ops |
| 7 | **unlink** | open, unlink, close | Link Ops |
| 8 | **symlink** | symlink, unlink | Link Ops |
| 9 | **chmod** | chmod | Attr Ops |
| 10 | **stat** | stat | Attr Ops |
| 11 | **fstatat** | fstatat | Attr Ops |
| 12 | **read** | open, read, close | Read Write |
| 13 | **write** | open, write, close | Read Write |
| 14 | **copy** | open, open, read, write, close, close | Read Write |

While there are 43 system calls (out of 382) for file accesses in Linux v5.3.0, we only evaluate a subset of them because they have more relevance to the hooking overhead we want to measure. For all 43 systems calls for file accesses, they fall into several categories: (1) file operations (e.g., open, stat); (2) directory operations (e.g., mkdir); (3) link operations (e.g., symlink); (4) basic file attributes (e.g., chown); (5) extended file attributes (e.g. setxatrr, getxattr, listxattr); (6) file descriptor manipulations (e.g. dup, fcntl); (7) file data read/write (e.g. pread, pwrite); and (8) auditing file events (e.g. inotify_init, inotify_add_watch). Those in category (5) and (8) are privileged operations for the root user and normal users have limited accessibility to them; those in category (6) do not trigger any hooks. Therefore, we do not measure the system calls in these three categories. For the rest categories, we test the representative system calls which are listed on Table 3. What's more, the set of the system calls we measure is exactly the same with those analyzed in [1] for POSIX standard. The 14 system calls in LMBench-Hook are enough to trigger the most-common filesystem hooks, which mediate shared Security Sensitive Objects (SSOs) (i.e., file, path, inode and dentry) [15]. When accessing a file, system calls in Table 3 invoke kernel handler. The kernel first accesses file and path after parsing the system call arguments. Then, dentry is further visited by referencing the field in file or path; inode can be accessed from the file d in dentry. Kernel APIs are called to manipulate these SSOs. To guarantee complete mediation, Linux performs policy enforcement to guard the access to these SSOs in these kernel APIs. Major security modules in Linux implement one or more hooks for each type of SSO. For example, SELinux (Linux v5.3.0) implements 10 hooks on file, 31 hooks on inode, 2 hooks on dentry. Different security module implements a different subset of hooks defined by LSM at their discretion.

We provide a summary of the benchmarks in LMBench-Hook in Table 3. The open benchmark measures the latency to open a file for reading and immediately closes it. The stat benchmark measures the latency to invoke stat system call on a file. Both open and stat include 11 sub-tests with directory depth from one to eight, a hard-link, a soft-link, and one non-existing directory test. The read/write/copy benchmark measures operations per second and the latency of each operation. For read/write benchmark, each read/write system call is one operation; a copy operation includes a read from the source file and a write to the destination file. In read/write/copy benchmark, we run the tests with various buffer sizes for system call read and write. For 0KB buffer size, system call overhead dominates the time of operation. Thus 0KB buffer is used for measure latency of read/write/copy. The hooking overhead consists of re-validating permissions for each read, write and copy. When buffer size increases (e.g.,1KB, 2KB and 4KB), memory copy cost become more significant to impact latency of system call read and write, so the hooking overhead becomes less noticeable. Thus, we do not test buffer size larger than 4 KB. rename and chmod test measure latency of invoking the corresponding system calls, each includes 5 sub-benchmarks with directory depth from one to five. openat, creat, unlink and symlink measure latency of operating on a particular file, with random filenames. mkdir and rmdir measure operation per second for 9437 distinct files, with directory depth of one and creating or removing a file is one operation.

To measure the overhead of the hooking without introducing authorization overhead, we use the securityfs interface exported by each security module to disable policy enforcement (e.g. policy checking). When policy enforcement is disabled, the functions for policy enforcement are bypassed while the hooks are still invoked.

## 4.2 Tunable Security Module for Latency Modeling

The overhead of LSM framework comes from two aspects, (1) hooking (e.g, security attributes manipulation, hook placement), (2) policy enforcement. The hooking overhead varies depending on the hook placement. Nevertheless, for a specific filesystem operation on a given security module, this hooking overhead can be treated as a constant. On the other hand, policy enforcement overhead may change even for the same security module. For example, the time to evaluate the rules of the policy against an access request may differ considerably for different policy configurations. Even the underlying data structures used for the policy store affect the efficiency of the enforcement of the policy. However, it is a complex task to understand how the variations in policy enforcement impacts the end-to-end performance of file operations. We try to approach this issue by studying how sensitive the end-to-end latency of a file operation is to the changes of the latency of policy enforcement. We assume there is no interaction between the effect of policy enforcement and hooking. Then we can describe the end-to-end latency of a file operation with a Multiple Linear Regression Model [20], for a given security module that enforces a fixed policy. We use $l_{\square\square}$ to denote the latency of a file operation, $l_{h\square\square\square}$ the latency from hooking mechanism, $t_{policy}$ the latency from policy enforcement, and $c$ for other constant cost. Then, we have Equation (1).

Table 4: Hooks in Policy Testing Module.

| No. | Name | Description |
| --- | --- | --- |
| 1 | *security_bprm_set_creds* | mediates loading of a file into a process (e.g., on exec), labeling the new process as described above. |
| 2 | *security_inode_alloc_security* | initialization of a new inode object, allocate memory space for security blob. |
| 3 | *security_inode_init_security* | mediates initialization of a new inode object, setting the label to that of the creating process. |
| 4 | *security_inode_setxattr* | mediates modification of the extended attributes of a file's inode. |
| 5 | *security_inode_getsecid* | mediates reading a file's the extended attributes of a file's inode (i.e. security tag). |
| 6 | *security_inode_create* | mediates the return of a newly created file to the process. |
| 7 | *security_file_permission* | mediates operations on a file descriptor, example operations include read, write, append. |
| 8 | *security_inode_permission* | mediates file open operations on the file's associated inode. |

$$t_{sys} = \beta_1 \times t_{hook} + \beta_2 \times t_{policy} + c \quad (1)$$

In Equation (1), $\beta_1$ and $\beta_2$ are partial regression coefficients. Our goal is to estimate $\beta_2$ to quantify how much impacts $t_{policy}$ can have on $t_{sys}$.

In this section, we describe the approach we use to estimate $\beta_2$. We develop a dummy security module to meet our goal and we name it the Tunable Security Module. The Tunable Security Module follows the design of SELinux [23, 24] and inherits from SELinux the hooks for mediating file accesses except the 8 hooks listed on Table 4. In these 8 hooks, security_inode_permission and security_file_permission are interfaces between hooking and authorization modules. Internally, hooks for file access, such as security_inode_unlink and security_inode_rename, call security_inode_permission and security_file_permission for "permission checking". The other 6 hooks are responsible for initialization and allocation of security blobs, getting/setting file attributes, getting attributes from user-space programs, and permission control on files/inodes. The Tunable Security Module implements security_inode_permission and security_file_permission as a busy-waiting function. The amount of time to busy-wait is passed from the user space through **securityfs**. We also implement the other 6 hooks according to their functionalities described in 4.

The Tunable Security Module has two execution stages: (1) initialization stage and (2) enforcement stage. For the initialization stage, the value (in $\mu$s) of the duration of the delay is passed to the kernel from user space. Enforcement stage handles authorization queries from the hooks for file accesses and imposes the delay on the queries and grants the permission.

The Tunable Module behaves as a normal security module expect policy enforcement. It implements the code to manipulate the security tags. The user space program can set security tags through setxattr and getxattr in string format (i.e., "trusted", "untrusted", "ignored"). Using the file-system's extended attributes, label strings are set as "trusted" , "untrusted" and "ignored" in the file's security.test attribute. They are translated into security xattr in inode->i_security, as an u32, with 0, 1, 2 stand for "trusted", "untrusted", and "ignored", respectively. If the executables (e.g. open.exe) or the test files (e.g. /test/1.txt) have no security tag, the default security tag, "ignored", is assigned. The Tunable Module implement its own labeling system. The kernel objects (e.g., processes and inodes) get their labels based on the labels of the files that are used to create them.

### 4.3 Limitations
This work has three limitations: (1) We consider Linux as a file based system. This work is focusing on testing file operations. Network and device driver are not considered, though studying hooking overhead for these subsystems are interesting topics. (2) As hooking and policy enforcement in LSM are in-memory operations, we are focusing on in-memory filesystem operations in this work. System call **mount** and **umount** are not considered. (3) The LSM framework supports various access control models. Each of the access control models has its own implementation of policy enforcement. Implementation of policy enforcement various and affects latency brought by policy enforcement. Furthermore latency introduced by policy enforcement is affected by how many rules and which rules users set. There are no standards on synthesizing the rules. Thus, instead of coming up with some imagined rule sets, we write a Tunable Security Module, which sets latency of policy enforcement to a certain value, and checks impact of policy enforcement on end-to-end performance.

## 5 EVALUATION RESULTS
This section presents hooking overhead evaluation and analysis for filesystems. We further make a few key observations before detailing them in Section 6.

**System Setup.** We conduct the tests on a 6th Generation Intel® Core™ i7-6500U 2.50 GHz processor with 2 cores. The machine also has 8 GB LPDDR3 1866MHz memory and a 512GB PCIe SSD for persistent storage. The tests are done with power cord on to avoid CPU frequency shifting. The machine has Ubuntu 18.04 LTS with Linux kernel v5.3.0. The tests are done on ext4 with default parameters. When evaluating SELinux, we set 512 as the maximum AVC entries in the cache.

**Evaluation Metrics** We use three evaluation metrics: (1) latency; (2) throughput; and (3) performance overhead. We report the latency or throughput (operations per second) for the 16 tests mentioned in Section 4.1. For each of the 16 tests, we run 300 times. Mean and variance of the data points are calculated and reported for tests. For test 1-5 and 8-13, we measure the latency for a single system call. For test 6 and 7, we measure throughput (i.e., operations per second) for mkdir and rmdir. For test 14-16, we pre-create a file with 100KB and then perform sequential read or write upon this file or sequentially copy this file to a new file. We just wrap

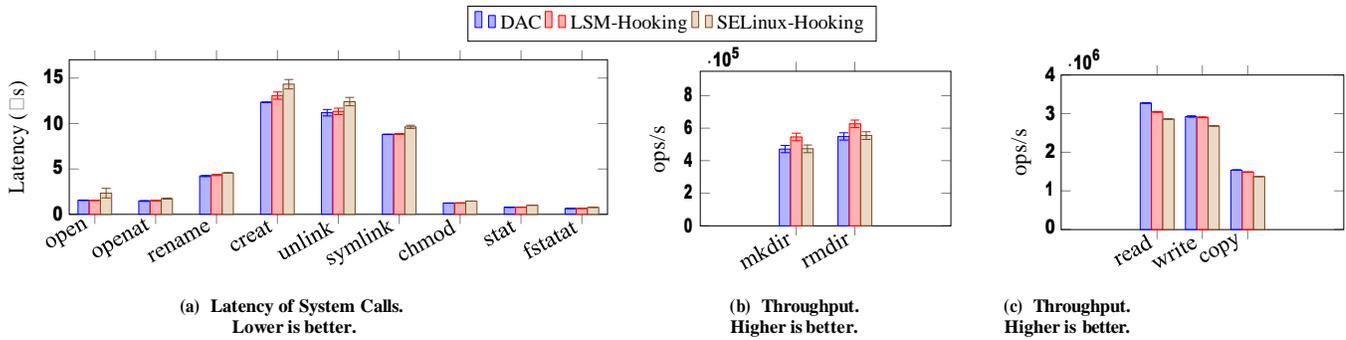

(a) Latency of System Calls. Lower is better.

(b) Throughput. Higher is better.

(c) Throughput. Higher is better.

**Figure 2: Performance Comparison of the Kernel without Hooks and with Capabilities or SELinux Hooks.** Default depth setting in LMBench2.5 for open and stat, mkdir/rmdir and others are tested with folders of one depth directory, read/write/copy with 0KB files. Overhead for open is small in absolute value (about 1 □□ s), however the absolute value is higher for low-end embedded systems [26].

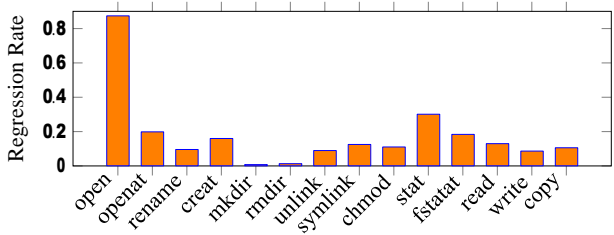

**Figure 3: Performance Overhead of SELinux. Smaller is better.** Performance drop of open (0.87) and stat (0.30) are higher than histori- cal evaluation.

around when tests reaches the end of the file. For these three tests, we first run the test 10 seconds to warm up the cache, and then run the test for 30 seconds. We measure the throughput (i.e., operations per second) for the second phase. To make a comparison of overhead of each syscall before and after hooking, we also report it with unmodified kernel v5.3.0 (LSM not enabled) as baseline, which is denoted as □□□□_□□□. The performance of the targeted testing's performance is denoted as □□□ _□□□□□. Regression rate (i.e., overhead) is calculated with Equation (2).

$$\mathit{Regression\ Rate} := \frac{|\mathit{test\_target} - \mathit{test\_base}|}{\mathit{test\_base}} \quad (2)$$

## 5.1 Historical Evaluation Revisited

We first revisit the hooking overhead of SELinux and LSM since previous evaluation [19, 24] was done 18 years ago. In [24], the authors evaluated performance overhead of capabilities security module and the baseline was unmodified Linux kernel (v2.5.15) without LSM; in SELinux [19], the authors evaluated performance impact of the SELinux and the baseline was unmodified Linux kernel (v2.4.2) with only DAC protection. The unmodified Linux v2.4.2 kernel includes capabilities, which is moved into LSM as its default security module since Linux v2.5.29. The tests in LSM [24] and SELinux [19] set a policy in authorization module. However, no labels are added to files as attributes, and system calls get short circuited once they enter policy enforcement. To re-measure the hooking overhead, we compare kernel v5.3.0 with three different configurations—with only DAC protection, with default LSM (only capabilities module) and with default SELinux module (i.e., SELinux stacked with capabilities module, auditing is enabled). For the latter two, policy enforcement of the hooks is disabled as mentioned in Section 4.1. The result is shown in Figure 2. Compared with pure DAC kernel, the hooking overhead of LSM is small for all the tests. Compared with the results in [24], LSM hooking is still efficient and does not cause tangible performance impact. However, SELinux hooking could cause large performance drop for open (87%) and stat (30%), overhead for open is small in absolute value (about 1 □□s), however the absolute value might be higher for low-end embedded systems [26]; the overhead for mkdir and rmdir is smaller than 2%; for the rest of the tests, the overhead ranges from 8% to 19%. We also report the performance overhead (regression rate) of SELinux hooking using Equation (2), see Figure 3. Compared with the results in [19], SELinux hooking cause more significant overhead; in [19], the overhead of SELinux was no larger than 28% for the tests we evaluate, smaller than what we observe.

*5.1.1 Hooking Overhead Comparison of LSM-based security mod- ules.* Different Linux distributions uses different LSM-based secu- rity modules, for example, Ubuntu has AppArmor turned on by default while Fedora has SELinux turned on by default. Different LSM-based security modules implement different subset of hooks. As is shown in Table 5 and Table 2, in Linux v5.3.0, capabilities module implements 767 LOC with 18 hooks, 4 of which are for file accesses. SELinux implements 21,266 LOC with 204 hooks, 59 of which are for file accesses. AppArmor implements 11,918 LOC, with 68 hooks, 24 of which are on file accesses. SMACK implements 5,369 LOC with 108 hooks. TOMOYO implements 28 hooks with 8,245 LOC. In this section, we evaluate how different LSM-based security modules perform. And as different LSM-based security modules uses different sets of hooks. We further investigate how different selections of hooks impact the performance by evaluating the hooking overhead of 5 existing security modules which are capabilities, SELinux, AppArmor, TOMOYO, SMACK. We are eval- uating hooking overhead, thus policy enforcement code is disabled when we compare LSM-based security modules. Different security modules impact benchmarks in different ways. Capabilities mod- ule impacts all benchmarks. SELinux module impacts open higher than openat. AppArmor impacts open, openat, rename, creat and especially mkdir and rmdir. TOMOYO with no integrity measurements added introduce tolerable performance overhead. SMACK has moderate impact on all benchmarks, except for mkdir, where it has significant low impact. The impact on file read is higher than file write, and the impact of file copy is between the two. We configure the kernel to have only one of the 5 modules to be active. The overhead introduced by the hooking of each security module is shown in Figure 6. For all the tests, the overhead of hooking for each individual module is within 15%. For creat, capabilities, SELinux and AppArmor have overhead slightly larger than 5%; for mkdir and rmdir, capabilities, SELinux and AppArmor have overhead ranging from 8% to 14%; for stat, capabilities causes 5.6% overhead; for read, the overhead of all modules ranges from 5.8% to 8%; for the other tests, overhead is smaller than 5% for all modules. We firstly use LMBench-Hook to collect frequencies of security hook executions for each benchmark. security_file_permission accounts for 99% of security hooks called by read, write and copy. security_inode_permission and other hooks on inode structure, such as security_inode_getattr and security_inode_follow_link, account for 99% of security hooks called by stat. inode related hooks accounts 60% and file related hooks account for 33% of security hooks called by open. In summary, security_file_permission and security_inode_permission dominates all 14 benchmarks' execution path. These 14 benchmarks, which include 59 sub-benchmarks, have to pass either security_file_permission or security_inode_permission, no matter if the syscall is a successful return or not. To our surprise, mkdir and rmdir are not associated with any hooks in Linux v5.3.

*5.1.2 Overhead of Module Stacking.* Starting from Linux version 5.3.0, users are given the flexibility to configure stacking order of security modules. In this section, we evaluate how stacking order of security modules impacts performance. We evaluated hooking overhead when the system has 2 active modules. We stack SELinux, AppArmor, SMACK and TOMOYO on top of capabilities, or vice versa. In total, we have 8 configurations. For each pair of security modules which are stacked together, we compare the overhead of different stacking orders. The result is shown in Figure 10. From a high level, the hooking overhead of two active modules is larger than when there is only one active module in many cases. For example, when SMACK is stacked before capabilities, the median of regression rate for all tests is 15.3%, while for SMACK and capabilities alone the respective median is 1.5% and 3.6%. Similar results also happen to the other three pairs of modules. In addition, we find that for capabilites and TOMOYO, the regression rate is larger than 100% for mkstemp, unlink, symlink, chmod, stat and fstatat.

*5.1.3 Overhead of Integrity Measurements.* This section evaluates the performance impact of Integrity Measurements in LSM. To analyze trade-offs of the combinations, we measure: How auditing, IMA and EVM impact hooking performance? Integrity module is a stack-able module. Integrity module's 12 hooks (v5.3.0) are embedded in general hooks. As is shown in Table 5, integrity module implements two categories of hooks, EVM hooks (5) and IMA hooks (7). EVM has 5 hooks on inode data structure. IMA has 2 hooks on inodes, 3 hooks on file, 1 hook on mmap and 1 hook on bprm. Two major modules (i.e., SELinux and SMACK) invoked all these hooks. We take SELinux as an example to evaluate performance downgrade brought by integrity module. We evaluate performance

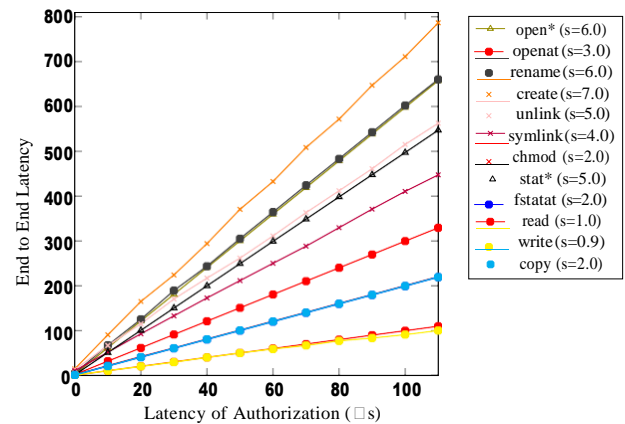

* open and stat is tested with default input in LMBench2.5, which is opening and stating file /usr/include/x86_64-linux-gnu/sys/types.h.

**Figure 4: End-to-End Latency of the Tests by Increasing the Latency of Policy Checking (left) and the Slopes Calculated with Linear Regression Method (right). The policy checking latency is much larger than the time spent on the rest parts of the benchmark (expect 0). Slope varies by tests. The $R^2$ values of linear regression is 0.999. The higher the slope, the more significant impact is. The slope reflects times authorization is invoked by a certain test.**

of DAC Linux and SELinux with audit on, and the results is shown in Figure 11. Both IMA and EVM introduces significant overhead across all benchmarks. EVM and IMA together brings overhead of 135% on chmod. mkdir, rmdir, fastat, read, copy, link and unlink gets non-tolerable (more than 50%) performance overheads.

## 5.2 Impact of Hook Placement on File Accesses

In previous sections, we do not take the performance impact of policy checking into consideration so that we can evaluate and compare the impact of hooking across different security modules. In this section, we evaluate the performance impact of policy checking on the end-to-end latency of file accesses by using a Tunable Security Module introduced in 4.2. Also, we conduct static analysis for hooks and their placements for different LSM-based security modules.

*5.2.1 Performance Analysis of Hook Placement on File Accesses.* In the experiment, we tune the latency of policy checking from 0 to 110 $\mu$s and measure the end-to-end latency of all the system calls we test, except for mkdir and rmdir. We plot the result in Figure 4. We observe that the relationship between policy checking latency and the end-to-end latency of system calls are nearly linear. We use linear regression method to calculate the linear coefficient of the data points and the results are shown on the right of Figure 4. This coefficient reflects times authorization is invoked with each benchmark. As shown in Figure 4, end to end latency and latency introduced by policy authorization (i.e., policy querying, parsing, processing and checking) are linearly proportional to one another with determine of 0.99. However, their impact factor on end to end latency (slope in Figure 4) differ. For example, slope of openat is 3.0, while rename is 6.0.

Furthermore, as is shown in Table 6, slope increases while directory depth increases. For open and stat, we change the input path and re-calculate the linear coefficient. As shown Table 6, we find for different paths, the linear coefficient increase as the number of components in the path increases. However, for the other tests, when we change the input path, the linear coefficient stay the same. This means the times authorization invoked vary by different paths for open and stat. However for other tests, the times authorization invoked is a constant value.

Table 5: Hook Placement of Security Modules (Linux version 5.3).

| Hook Num (by category) | Capability | SELinux | AppArmor | SMACK | TOMOYO | YAMA | EVM | IMA** |
|---|---|---|---|---|---|---|---|---|
| **inode** | 3 | **31** | 1 | **22** | 1 | 0 | 5 | 2 |
| **dentry** | 0 | 2 | 0 | 1 | 0 | 0 | 0 | 0 |
| **file** | 0 | **10** | 7 | **8** | 3 | 0 | 0 | **\*\*3** |
| **superblock** | 0 | **13** | 3 | 6 | 3 | 0 | 0 | 0 |
| **mmap** | 2 | 2 | 1 | 2 | 0 | 0 | 0 | **\*\*1** |
| **path** | 0 | 0 | **10** | 0 | **11** | 0 | 0 | 0 |
| **bprm** | 1 | 3 | 3 | 1 | 2 | 0 | 0 | 1 |
| **task** | *5* | *15* | *5* | *12* | *2* | *2* | 0 | 0 |
| **proc** | 0 | *2* | 0 | *2* | 0 | 0 | 0 | 0 |
| **ptrace** | 2 | 2 | 0 | 2 | 0 | 2 | 0 | 0 |
| **cap** | *3* | *3* | *2* | 0 | 0 | 0 | 0 | 0 |
| **seclabel** | 0 | 3 | 0 | 3 | 0 | 0 | 0 | 0 |
| **cred** | 0 | 3 | 4 | 5 | 1 | 0 | 0 | 0 |
| **audit** | 0 | 4 | 4 | 3 | 0 | 0 | 0 | 0 |
| **Total(File Accessing*)** | 4 | 59 | 24 | 38 | 20 | 0 | 5 | 6 |
| **Total Num** | 18 | 204 | 68 | 108 | 28 | 4 | 5 | 7 |

\* We consider hooks on inode, dentry, file, superblock, path, bprm are file accessing hooks. \*\* IMA has three file related hooks (i.e. ima_file_mmap, ima_read_file, ima_post_read_file), and one mmap related hook. This mmap related hook only performs on files (i.e. ima_file_mmap), however not general mmap.

Table 6: Directory depth impacts latency of open and stat (LMBench-Hook). The first column is the path we use in the open and stat tests. The last two columns report the slope of linear model we build, with r-square value of 0.999. The linear model reveals that there is positive correlation between the latency of policy enforcement and the latency of end-to-end tests. The slope reflects times authorization is invoked by a certain test. The higher the slope, the more impact.

| Path | open | stat |
|---|---|---|
| *AA* | 2.0 | 1.0 |
| *AA/BB* | 3.0 | 2.0 |
| *AA/BB/CC/DD* | 5.0 | 4.0 |
| *AA/BB/CC/DD/EE/FF/GG/HH* | 8.9 | 7.9 |
| *AA/../HH* | 4.0 | 3.0 |
| *XX/YY/../../AA/BB/../../HH* | 9.9 | 8.9 |

*5.2.2 Static Analysis of Hook Placement on File Accesses.* We perform static analysis on call graphs for understanding worst case scenarios of hook invocations in execution of VFS syscalls, as is shown in Table 7.

We conduct our static analysis based on the fact that all permissions could be categorized into read/write permissions either on files, file descriptors or on files' containing directories. And one authorization hook could perform read, write or read-and-write permission checks. Call graph analysis explains the maximum amount of hooks invoked by syscalls, among which, some are not invoked based on flags passed to LSM interfaces.

**Static analysis on call graphs.** In our enhanced static analysis on call graphs, we generate call graphs from call strings analysis, then enhance the call graph with call back routine edges. For example, in the sample program shown below, we firstly generate the three edges from call string analysis, main→foo1, main→foo2, foo1→fun. Since there is a call back routine after foo1() function is executed, we add another enhanced edge to the graph, foo1→foo2. Then the generated graph is main→foo1, main→foo2, foo1→fun, foo1→foo2. As all statements are executed before foo2 is called, a call back routine should be reflected in the graph. If there is a hook in foo1() function, the hook should be considered as already invocated for foo2() function as well. Through adding edge foo1→foo2, we better explains statements executed before foo2().

```
1       foo1(){
2           fun();
3       }
4
5       main(){
6           foo1();
7           foo2();
8       }
```

**Static analysis results.** Minimum hooking is reasoned from POSIX's definition of syscalls. We conservatively assume file descriptors could be read/written when their associated files has read/write permissions, which is different form LSM. In LSM, file descriptors, which point to entries in the kernel's global file table, are not associated with any permission checks. However, according to POSIX, the kernel is supposed to return a file descriptor, only after a process makes a successful request to open a file. And opening a file requires read permission. In summary, file descriptors should hold the same permission as their associated files. For example, a file requires read permission to perform syscall *stat*. The details of reasoning is as follows. *open* searches, opens and possibly create a file, and read permission is required for the file's containing directories and the file itself. Write permission is required for the file's direct containing directory, if *open* is flagged with CREAT. Meanwhile, *openat* syscall opens by a file descriptor, read permission is required for the file itself. *close* closes a file descriptor, no permission is required during this process. *rename*, when both parent folders exist, and parent folders are different, requires read permission and write permissions on the two files (newly created one, and the original one), and the two associated direct containing directories. *sendfile* requires read permission on one file and write to another file. *read/write/chmod* requires read permission on the file itself, write permission on the file itself, and write permission on the file descriptor (i.e. metadata) respectively. *mkdir/rmdir* requires write permissions on files' containing directories. *link/symlink/unlink* requires read permission for searching (execution permission, i.e. read permission) on its containing directories, and *link/symlink* also requires write permission on the file. *stat* obtains file and related filesystem status named by the pathname parameter. It requires read permission for the named file's file descriptor. Also, directories listed in the pathname, which leads to the file, must be searchable. Thus, read permission is required for its containing directories. Different LSM-based security modules implements different sets of hooks for permission authorization. Some security modules do not meet complete mediation requirement on call graphs of syscalls. For example, AppArmor and TOMOYO are path based permission authorization, when creating files, they do not need to request write permission for the new files' containing directories. SMACK and TOMOYO do not implement *file_permission* hooks on *sendfile/read/write*, and do not support security on above syscalls. While some security modules over-worked the permission authorization with duplicated policy checks. For example, SELinux and SMACK implements 7 hooks (5 authorization hooks in form of *security_inode_permission* and 2 authorization hooks in form of *security_inode_rename*) on permission authorization, while only 4 authorization hooks are required.

## 6 ANALYSIS OF RESULTS

This section discusses the performance overhead evaluated in Section 5. We also present some insights for optimizing LSM. For each root cause of the performance downgrade, we first review the background of the change before analyzing its performance impact.

**Hardware Evolution Introduce New Bottlenecks.** As mentioned in Section 5, performance overhead of LSM hooks is different to what reported in previous work [38]. We see much larger performance degradation (from 28% to 87%) in terms of latency for open system call when SELinux is enabled. Performance overhead of LSM framework. Bottleneck changes from I/O bottleneck to computation bottleneck. Over the past decade, storage and memory becomes more powerful, while CPUs are down for Moore's Law. Hooking gets more complex, from 122 hooks (Linux v2.6) to 224 hooks (Linux v5.3), from 43 lines per hook (Capability) to 104 lines per hook (SELinux) and 295 lines per hook (TOMOYO), which introduces increased computation. These introduces new bottlenecks for LSM's performance.

**Performance Impact of Stacking Order.** Starting from Linux v5, LSM adds flexible stacking order, to enable container and host for flexible security modules. For example, you could use SELinux in containers, while AppArmor on host [6] As we see in Section 5, stacking order matters to performance. In module stacking, the checking order follows a white-list based approach. We use an example to illustrate how the performance impact varies for different stacking orders.

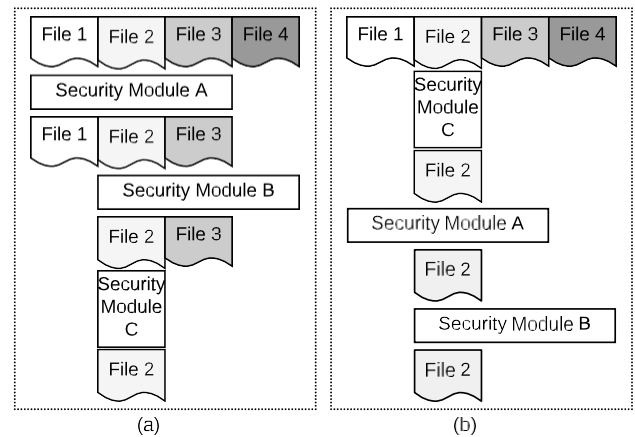

**Figure 5: White-list Based Module Stacking. Stacking order matters to performance.**

We use an example to illustrate how the performance impact varies for different stacking orders. As is shown in Figure 5 (a), security module A grants access to file 1, 2 and 3, security module B file 2, 3 and 4, and security module C file 2. If we configure the stacking order as CONFIG_LSM="A,B,C", then file 1-4 will be checked by security module A first. As security module A allows file 1-3, they will be further checked by security module B. Similarly, as security module B allows file 2-4, only file 2 and 3 will pass module B and be checked by module C. As security module C allows file 2, only the access to file 2 can be granted. In this process, 4 files are checked by module A, 3 files are checked by module B and 2 files are checked by module C. If we switch the stacking order as CONFIG_LSM="C,A,B", as is shown in Figure 5 (b), all 4 files will be checked by module C first. Module C only allows file 2. Thus, only file 2 will be checked by security module A and B. In this example, the second stacking order costs less time. If the latency introduced by each security module checking is one unit of time, as a result, stacking order depicted in Figure 5 (a) consumes three units of time more than that is consumed by Figure 5 (b). In summary, by changing the security module stacking order, the latency introduced by information flow authorization differs.

For example, we test and compare CONFIG_LSM = "capability, selinux, apparmor " and CONFIG_LSM = "capability, apparmor, selinux ", the latency of open and stat diffs more than 10% between the two settings. SELinux set white-listing on special files as in "proc", while apparmor set white-listing on special file types. A second reason why different stacking orders cause different performance overhead is that some security modules implement their own caching mechanism whereas others do not. For example, SELinux implements Access Vector Cache (AVC) while TOMOYO lacks of implementing any cache mechanism. Due to the effect of where the caching is layered, the order of module stacking can impact the performance.

Table 7: Hook Placement by Syscall (Linux version 5.3).

| Syscall Name | Similar Syscall | Min Hook | LSM Interface | SELinux | AppArmor | SMACK | TOMOYO |
|---|---|---|---|---|---|---|---|
| open | open | 1*dir depth | security_inode_permission | 3*2*dir depth | 0 | 3*2*dir depth | 0 |
| | | | security_file_open | 3*2*dir depth | 3*2*dir depth | 3*2*dir depth | 3*2*dir depth |
| openat | openat | 1 | security_inode_permission | 3*2 | 0 | 3*2 | 0 |
| | | | security_file_open | 3*2 | 3*2 | 3*2 | 3*2 |
| close | close | 0 | N/A | 0 | 0 | 0 | 0 |
| creat | creat | 1*dir depth | security_inode_permission | 1*dir depth | 0 | 1*dir depth | 0 |
| rename* | rename, renameat, renameat2 | 4 | security_inode_rename | 2 | 0 | 2 | 0 |
| | | | security_path_name | 0 | 1 | 0 | 1 |
| | | | security_inode_permission | 5 | 0 | 5 | 0 |
| sendfile | sendfile, sendfile64 | 2 | security_file_permission | 2 | 2 | 0 | 0 |
| read | read, readv, pread, preadv | 1 | security_file_permission | 1 | 1 | 0 | 0 |
| write | write, writev, pwrite, pwritev | 1 | security_file_permission | 1 | 1 | 0 | 0 |
| mkdir | mkdir, mkdirat | 1* dir depth | security_path_mkdir | 0 | 1*dir depth | 0 | 1*dir depth |
| | | | security_inode_mkdir | 1*dir depth | 0 | 1*dir depth | 0 |
| | | | security_inode_permission | 1*dir depth | 0 | 1*dir depth | 0 |
| rmdir | rmdir | 1* dir depth | security_path_rmdir | 0 | 1*dir depth | 0 | 1*dir depth |
| | | | security_inode_rmdir | 1*dir depth | 0 | 1*dir depth | 0 |
| | | | security_inode_permission | 1*dir depth | 0 | 1*dir depth | 0 |
| symlink | symlink, symlinkat | 1* dir depth | security_path_symlink | 0 | 1*dir depth | 0 | 1*dir depth |
| | | | security_inode_symlink | 1*dir depth | 0 | 1*dir depth | 0 |
| | | | security_inode_permission | 1*dir depth | 0 | 1*dir depth | 0 |
| unlink | unlink, unlinkat | 1* dir depth | security_path_unlink | 0 | 1*dir depth | 0 | 1*dir depth |
| | | | security_inode_unlink | 1*dir depth | 0 | 1*dir depth | 0 |
| | | | security_inode_permission | 1*dir depth | 0 | 1*dir depth | 0 |
| chmod | chmod, fchmodat | 1 | security_path_chmod | 0 | 1*dir depth | 0 | 1*dir depth |
| | | | security_inode_permission | 1*dir depth | 0 | 1*dir depth | 0 |
| | | | security_inode_setattr | 1 | 0 | 1 | 0 |
| fchmod | fchmod | 1 | security_path_chmod | 0 | 1 | 0 | 1 |
| | | | security_inode_permission | 1 | 0 | 1 | 0 |
| | | | security_inode_setattr | 1 | 0 | 1 | 0 |
| stat | stat, fstatat, lstat | 1*dir depth | security_inode_getattr | 1 | 1 | 1 | 1 |
| | | | security_inode_permission | 1*dir depth | 0 | 0 | 1*dir depth |

*rename, for the situation that both parent folders exists, and are two different parent folders.

**Repetitive Permission Check for Directories.** For Table 6, all tests access the same number of files, and the only difference among these tests is the input path. We can notice that the slope calculated for each test is proportional to the count of components in the input path. In the test, we measure the end-to-end latency of the system calls which is consisted of two parts, the latency of policy checking and the latency of other parts during the execution path of the system call. The second part is constant in this experiment. We increase the latency of policy checking from 0 to 110 $\Box$s with 10$\Box$s interval. The physical meaning of the slope is the number of authorization queries the system call makes. Since by executing the same test, number of files to be accessed does not change, we could conclude that the same authorization query, which is defined by a tuple (subject label, object label, file, operation), is performed multiple times for accessing directories.

The same open/stat test program accesses the same parent directory, and this is performed multiple times for the test files.

Redundancy is observed in security_inode_permission and security_file_permission in directory accessing of open and stat. This redundancy is observed within one system call instance, specifically on system call open and stat. These two system calls performs permission checks for path lookup, and open or stat on particular files in a specific directory, such as /xxx/1.txt and /xxx/2.txt. This consecutive access to the files in the same directory causes redundant hook invocations. This redundancy may be considered within one syscall instance (specific call to open) or across multiple open syscalls

As mentioned in Section 4.2, in each test the system call is executed 300 times consecutively. We can infer that all 300 system call queries need to go through the same permission check for each component in the file path even though they are visiting the same file in the same directory. This finding implies that it would be beneficial to cache the permission check results for directories when a file underneath it is visited. With this cache, future accesses to the files in the same directory can spend less time doing permission checks for the parent directory.

**Policy Enforcement of LSM-based security modules for LMBench-Hook.** As shown in Figure 4, these tests are insensitive to the count of components in the path names. For example, the openat test performs 3 permission checks and rename 6, for all types of path names according to our analysis. One test might go through LSM permission check several times. A file accessing related system call will result in an execution path, the execution path covers 3 layers. The 3 layers in the execution path have control dependency among them. The control dependency is the root cause for redundant checks. In the first step, the path from the user space is looked up in the kernel, a kernel object *path* and *file* are created for this program. Next, an dentry object is created by referencing the field in the path or file object. Kernel-level file accessing API (i.e., inode_operations) handles operations on inode and dentry. Together with mixed kernel-driver-based APIs, such as ksys operations, they look up inode and dentry and map them to superblock. From there, driver-based file accessing APIs (i.e., super_operations) handles operations on superblock and reflects the operations on storage. Linux performs authorization on all SSOs, thus hooks are placed on all 3 layers. Major modules in Linux implement at least one hook on each layer. For example, SELinux (Linux v5.3) implements 10 hooks on files, 31 hooks on inode, 2 hooks on dentry, 13 hooks on superblock. For the openat test, the process first needs to check execute permission of the parent directory so that the target file can be looked up in it. After looking up the parent directory, a file, dentry, and inode object are created for the target file. More permission checks on file or inode are needed before granting access to the file. Specifically, security_file_permission and security_inode_permission are invoked for the target file. Both checks are needed, as while one process is "lookup" a pathname, another process might make changes that affect the file. Additionally, security_file_fcntl is introduced by preparation stage of the openat test in LMBench2.5, which invokes permission as well. The existence of symbolic link is be a plausible reason for enforcing both permission check for file and inode objects. The hooks for file and inode permission check are security_file_permission and security_inode_permission, respectively. Symbolic link seems to be a plausible reason for enforcing both permission check for file and inode object. However, symbolic link is a special file with its own inode, different to the file or directory it points to. An interesting question is that ideally, a particular syscall passes only one time of policy authorization for accessing a certain file. Logical redundancy sits in benchmarks where slopes are larger than 1. The slope should be equal or bigger than one to full-fill complete mediation requirements. As shown in Figure 4, the higher the slope is, the more policy authorization it passes for the particular test. For example, openat passes policy authorization for 3 times, while rename passes policy authorization 6 times. Thus, rename is more sensitive than openat, in terms of latency of policy authorization. The more sensitive to policy authorization, the more non-stable hook placement it is, the worse the implementation is. Their placement locations should be carefully adjusted to avoid redundant checking for real world applications, such as (1) redundant hook invocations introduced by module stacking; (2) redundant hook invocations introduced by hook placement; and (3) redundant hook invocations introduced by workload pattern.

**Performance-Oriented Hook Placement.** We further investigate the impact of the count of hooks on performance. Intuitively, the more the number of hooks is, the larger the overhead is. As is shown in Table 5, SELinux has 31 out of 204 hooks for inode, and SMACK has 22 out of 108 hooks for inode; AppArmor 10 out of 68 hooks for path and TOMOYO has 11 out of 28 hooks for path. While TOMOYO has only 28 hooks, which is the smallest of all, its performance overhead is highest when stacked with capabilities module (which is enabled by default in Linux), as is shown in Figure 10. The computational complexity of the implementation of the hooks is another factor we need to consider to explain the hooking overhead. A simple metric computation complexity of hooks is the number of lines of code in the implementation. Also, lines of code for each hooks varies by hook implementations, see Table 2 summarizes the number of lines of code for each hook. Capabilities and TOMOYO has 295 LOC per hook and 50 LOC per hook, respectively. SELinux and SMACK both implement security_file_permission. However, the number of lines of code they use for implementing this hook are different. Latency introduced by each hook implementation matters more to performance of user-space programs (see Section 5), than that of hook numbers.

Previous hook placement works [7, 10, 16, 25, 26] try to minimize the count of hooks, not performance (i.e. hook invocations). Alternatively, hook placement algorithms could take performance as the objective.

## 7 RELATED WORKS

This section discusses two categories of related prior work: evaluation and analysis of Linux Security Modules and benchmarks on file accessing.

**Evaluation and Analysis of LSM.** LSM was first introduced by Morris et al. [24] in 2002 as a general framework to provide strong system security for Linux kernel. It shows that performance overhead caused by LSM is tolerable, less than 8%, with a capabilities module compared with an unmodified Linux kernel with built-in capabilities support. However, the industry has made significant advancement to the hardware of computer systems since then. This makes the evaluation results of [24] less relevant now. In our work, the evaluation is done on a computer with modern hardware; especially, its storage system is equipped with an NVME device.

Previous evaluation of hooking are done for Asbestos [8, 40], HiStar [44], Flume [18] and Laminar JVMs [30, 32]. However they are not evaluating main stream works that are merged into Linux. Since the advent of LSM, various mandatory access control policies, such as SELinux [38], AppArmor [2], TOMOYO [13] and Smack [34], have been implemented for it in Linux kernel. Though these work provide thorough implementation details under the LSM framework, the performance impact of them is not evaluated. Recent literates on evaluation of policies are based on simulation results [27], however not on real world systems. Recent work, PeX [45] presents effectiveness of hooks through a static permission check analysis framework for Linux kernel. It uses a novel and scalable indirect call analysis technique to generate the inter-procedural control flow graph and automatically identifies all permission checks and infers the mappings between permission checks and privileged functions. However, these works lack comprehensive evaluation in efficiency. Moreover, our work also made a comparative evaluation among security modules.

**Evaluation and Analysis of Other MAC Policies.** Various security mechanisms for enforcing secure information flow have are proposed over the years. Policies like type enforcement (TE), role based access control (RBAC), multilevel security (MLS), Biba [4], Bell-LaPadula [3], Clark Wilson, CW-Lite [16], LOMAC [9], etc. are implemented in Linux v5.3. Jaeger et al. [17] presented an approach for analyzing the integrity protection in the SELinux example policy. Zanin and Mancini presented a formal model called SELAC [43] for analyzing an arbitrary security policy configuration for the SELinux system. Rueda et al. [14] provided a formal semantics for the MLS policy in the SELinux OS. However, all these work lack performance evaluation results for policies. Recent work by Ronit Nath et al. [27] evaluates efficiency of policies in attribute-based access control (ABAC). However, their experiments are based on simulation, which is not on real-world systems.

**Benchmarks on File Operations.** As stated in [28, 41], when researchers reason about completeness and correctness of POSIX standards in file-systems, they analyze 14 system calls. In this paper, this method is followed. Previous standard filesystem benchmarks are using Intel lkp-tests suite [5] and previous papers [24, 31, 42]: (1) filebench [21], (2) lmbench (2.5 and [22]) (3) FS-Mark [29] and (4) unix-bench [39]. lmbench3 [22] adds scalability test to lmbench2.5 [22], however it misses chmod/rename etc., which are essential for security performance tests. For common functions (i.e., read, write, open, close, stat etc.), lmbech2.5 and lmbench3 [22] uses exactly the same function and implementation. FS-Mark includes file-size sensitive tests. It is focusing on various of file-sizes, in security test, in memory tests are needed. Thus the smaller the files, the better. unix-bench [39] adds file-copy, file-read, file-write. filebench [21] adds readwholefile (open once, then read several times, then close once), writewholefile (open once, then write several times, then close once), appendfile (open, stat, set offset, write) etc. for large file processing. Also, security tests for open, close and read should be timed separately. We are inspired by these four benchmarks. Our benchmark times individual syscall latency, not by benchmark, and adds directory depth tests and file size tests.

## 8 CONCLUSION

In this work, we evaluate the hooking overhead of Linux Security Modules. We find while the hooking overhead for the LSM framework is similar to what was reported in the previous evaluation, the hooking overhead of SELinux is much alarming for certain system calls (i.e., open and stat). We also evaluate and compare the hooking overhead of five security modules, capabilities, SELinux, AppArmor, SMACK, and TOMOYO. The performance impact of module stacking is also investigated. In general, stacking one module before another causes larger hooking overhead. We also find stacking order can impact performance. Moreover, the impact of the latency of policy enforcement of a security module on the end-to-end latency of file accesses is studied. In summary, this work provides comprehensive evaluation and analytic results for today's LSM and LSM-based security modules (on Ubuntu 18.04 with Linux v5.3.0).

## 9 PERFORMANCE OVERHEAD OF LSM-BASED SECURITY MODULES

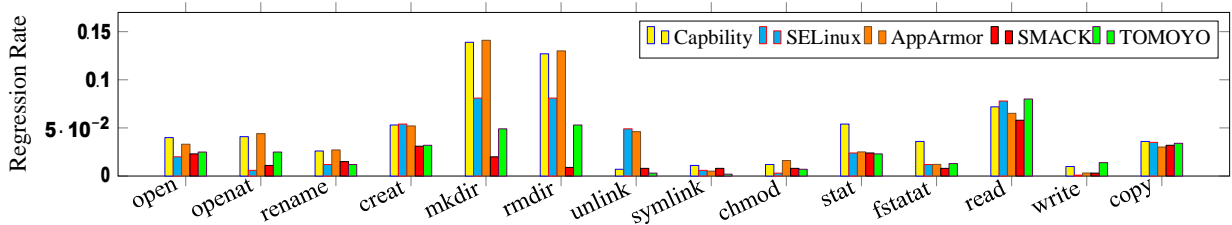

Figure 6: Performance Overhead of LSM-based Security Modules. Lower is better. Tested with directory depth of one.

## 10 PERFORMANCE OVERHEAD OF STACKING ORDER

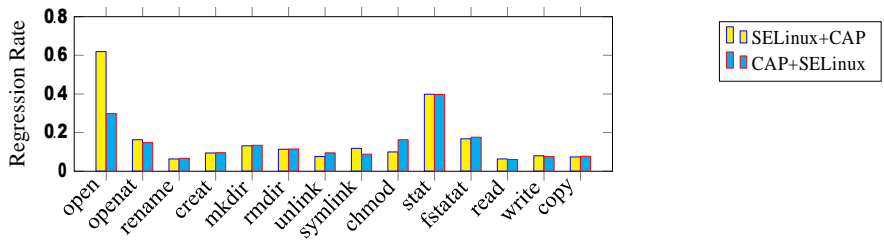

Figure 7: Overhead of Different Stacking Orders of SELinux and Capability module. In regression rate, the lower the better. Tested with directory depth of one. SELinux caches permission check results for each request for faster repetitive requests.

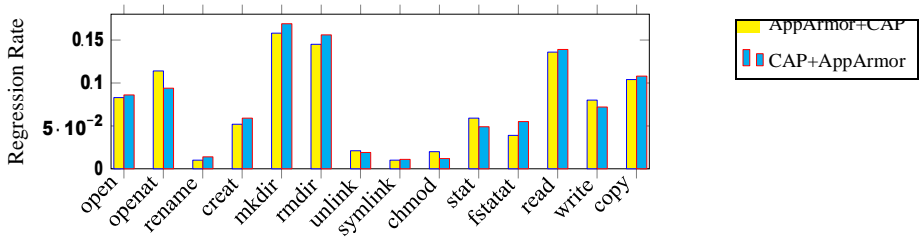

Figure 8: Overhead of Different Stacking Orders of AppArmor and Capability module. In regression rate, the lower the better. Tested with directory depth of one. AppArmor caches policies, and make it as a binary format when conducting checking to save run-time check execution time.

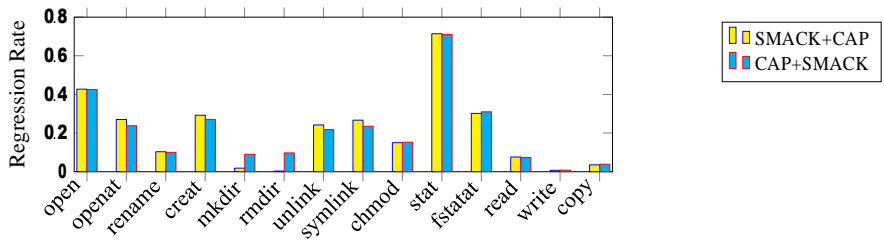

Figure 9: Overhead of Different Stacking Orders of SMACK and Capability module. In regression rate, the lower the better. Tested with direc- tory depth of one.

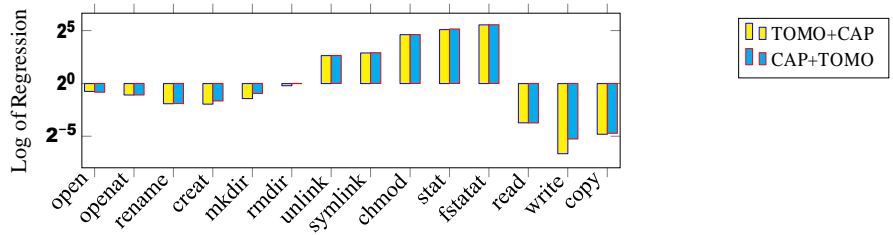

**Figure 10: Overhead of Different Stacking Orders of TOMOYO and Capability module. In regression rate, the lower the better. Tested with directory depth of one. TOMOYO shows significant negative performance overhead, as there lacks cache for accesses.**

## 11 PERFORMANCE OVERHEAD OF INTEGRITY MEASUREMENTS

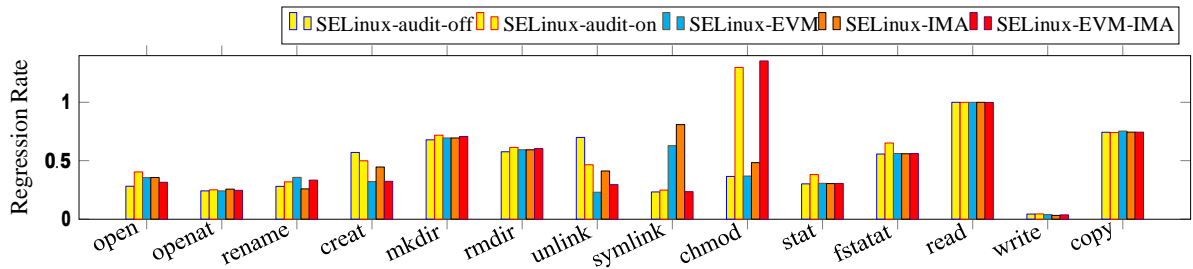

**Figure 11: Overhead Introduced by Integrity Measurements in LSM-based Security Modules, in regression rate, taking SELinux as an example. Tested with directory depth of one.**